\documentclass[a4paper,12pt]{article}

\usepackage[psamsfonts]{amssymb}
\usepackage{latexsym}
\usepackage{multirow}
\usepackage{color}
\usepackage{bm}
\usepackage{amsmath,amsthm}
\usepackage{caption}
\usepackage{setspace} 
\usepackage{authblk}   

\usepackage[draft]{changes}

\usepackage[authoryear]{natbib} 

\usepackage{graphicx}

\def\bSig\mathbf{\Sigma}

\newtheorem{theorem}{Proposition}

\usepackage{array}
\newcolumntype{H}{>{\setbox0=\hbox\bgroup}c<{\egroup}@{}}
\newcolumntype{Z}{>{\setbox0=\hbox\bgroup}c<{\egroup}@{\hspace*{-\tabcolsep}}}

\def\btheta{{\bvec \theta}}

\def\bvec#1{\mbox{\boldmath $#1$}}

\def\rI{\mathrm{\bf I}}

\title{\huge {N$_c$-mixture occupancy model}}

\author[1]{Huu-Dinh Huynh}
\author[2]{Wen-Han Hwang}

\affil[1]{Institute of Statistics, National Chung Hsing University, Taichung, Taiwan}
\affil[2]{Institute of Statistics, National Tsing Hua University, Hsinchu, Taiwan \authorcr\tt
wenhan@stat.nthu.edu.tw}

\date{}

\begin{document}
\maketitle

\begin{abstract}
A class of occupancy models for detection/non-detection data is proposed to relax the closure assumption of N$-$mixture models. 
We introduce a community parameter $c$, ranging from $0$ to $1$, which characterizes a certain portion of individuals being fixed across multiple visits. 
 As a result, when $c$ equals $1$, the model reduces  to the N$-$mixture model; this reduced model is shown to overestimate abundance when the closure assumption is not fully satisfied.
 Additionally, by including a zero-inflated component, the proposed model can bridge the standard occupancy model ($c=0$) and the zero-inflated N$-$mixture model ($c=1$). We then study the behavior of the estimators for the two extreme models as $c$ varies from $0$ to $1$. 
 An interesting finding is that the zero-inflated N$-$mixture model can  consistently  estimate the zero-inflated probability (occupancy) as $c$ approaches $0$, but the bias can be positive, negative, or unbiased when $c>0$ depending on other parameters.  We also demonstrate these results through simulation studies and data analysis.\\
 
\end{abstract}


\maketitle
\newpage
\section{Introduction}
Estimating the occupancy and abundance of species in a specific region, despite imperfect detection, is a crucial problem in ecology conservation and management. In reality, a species may exist at a survey site but go undetected due to limitations in the survey method or timing. Zero counts in a site sampling survey can be caused by the species not being present or by detection errors. If this issue is not addressed, the occupancy rate will be underestimated. Site occupancy models \citep{mackenzie2002}, based on multiple-visit detection/non-detection (occurrence, presence/absence)  data, can estimate the occurrence rate by accounting for detection errors. These models, which can be temporal or spatial replication surveys, are cost-effective as they do not require individual identification or marking. They are widely used in species distribution modeling, with various extensions such as multi-season open models, multi-species models, dynamic models, and spatial-temporal models \citep{mackenzie2017,hogg2021,mackenzie2009,johnson2013} having been developed. This study focuses on single-species, single-season occupancy models.

In the standard occupancy model \citep{mackenzie2002}, detection probability and species abundance are confounded and cannot be distinguished from one another. To overcome this limitation, \citet{royle2003,royle2004} introduced N$-$mixture occupancy models that allow for the separation of detection probability and species abundance. These models enable the estimation of species abundance through multiple-visit occurrence or count data. N$-$mixture models have received a lot of attention in the literature \citep{haines2016a,haines2016b,joseph2009,gomez2018} as they can estimate population size like capture-recapture models, but without the need for capturing and marking individuals. However, the performance of these models depends on the assumptions made \citep{dennis2015,link2018,barker2018}. Therefore, our goal is to improve these models by relaxing the closure model assumption, which is often violated even in single-season surveys \citep{kendall2013,otto2013}.

The closure assumption states that the number of individuals at a site remains constant during multiple visits. Surveys are often conducted through temporal or spatial replication, and sometimes with the use of multiple detectors or a combination of these methods \citep{mackenzie2017,KW09}.
Temporal replication involves surveying the same sites at different times, while spatial replication involves selecting random sampling units within a larger area at a single site. While this closure assumption is reasonable for surveys conducted in the same locations over a short period, it may not be appropriate for highly mobile species \citep{hayes2015}.  We also  note that the study in \cite{royle2004} is an example of temporal replication, despite the paper's title referencing spatial replication to indicate the distribution of sites.

The inference of the N$-$mixture occupancy model has been shown to produce biased point estimates and incorrect interval estimates when the closure assumption is violated. While a few studies have highlighted these effects \citep{denes2015,duarte2018,dail,ke2022}, most evidence is derived from simulation studies. To the best of our knowledge, there is a lack of theoretical results that explain the behavior of estimation bias under the N$-$mixture model. However, we have recently provided theoretical evidence under the proposed N$_c-$mixture model, which is an extension of the N$-$mixture model. It is important to note that \cite{dail} also proposes a class of generalized N$-$mixture models, which allows for the immigration and emigration of species populations and estimates year-to-year immigration and emigration rates, providing valuable insights for conservation managers. However, this model is a multi-season open population model that goes in a different direction than our extension.

\begin{figure}[ht]
\begin{center}
\includegraphics[width=6.3in]{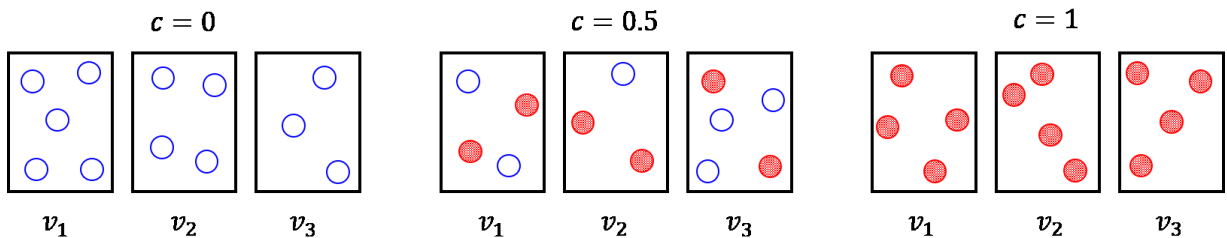}
\end{center}
\caption{
An illustration of the N$_c-$mixture model. In a triple-visit survey of a site, some individuals are present in the site for each of the three visits (denoted as $v_1$, $v_2$, and $v_3$). The figure illustrates the situations corresponding to the community parameter $c=0$ (left), $ c=0.5$ (middle), and $c=1$ (right). The colored circles represent those individuals who are residents (or, equivalently, are fixed) in the site during the three visits. For $c=0$, the number of individuals can differ from visit to visit, mimicking the scenario of spatial replication. For $c=1$, the number of individuals is constant, and the closure assumption of the N$-$mixture model is satisfied. For $c=0.5$, around half of the individuals were residents during the survey.
}\label{FG1}
\end{figure}

The N$_c-$mixture model is designed to create a framework that can unify both temporal and spatial replicating surveys. For example, consider a triple-visit survey conducted at a site where $N_1$, $N_2$, and $N_3$ represent the number of observable individuals during each visit. These variables are assumed to be identically distributed random variables. In the case of temporal replication, where the surveys are conducted in a short period, the three $N_j$ variables can be considered equal and meet the closure assumption of the N$-$mixture model. In contrast, for spatial replication, the $N_j$ variables are treated as independent. To account for both scenarios, we decompose the $N_j$ variables into two components: $N_j=K+M_j$ for $j=1, 2, 3$, where $K$ represents the number of common individuals during the triple-visit survey, and $M_j$ represents the number of non-common individuals. We assume that $K$ and $M_j$ are independent, with $E(K)=cE(N_j)$ for some $0\le c \le 1$. This parameter $c$, referred to as the \textit{community parameter}, indicates the proportion of individuals who are residents and remain fixed during the triple-visit. It also allows us to easily let $K$ degenerate to $0$ if $c=0$, or let $M_j$ degenerate to $0$ if $c=1$. Figure \ref{FG1} provides an illustration of this decomposition.

 Under the N$_c-$mixture model, we are able to demonstrate that  if the community parameter $c$ is incorrectly specified as $1$ (i.e., the N$-$mixture model is used instead), the mean abundance will be overestimated. Additionally, our results indicate that the bias increases as the community parameter $c$ decreases, reaching infinity as $c$ approaches $0$.

We propose an extension to the N$_c-$mixture model by incorporating a zero-inflated component. This allows us to bridge the standard occupancy model when $c=0$ \citep{mackenzie2002} and the zero-inflated N$-$mixture model when $c=1$ \citep{haines2016a}. We then investigate the behavior of estimators for these two extreme models as the community parameter $c$ ranges from $0$ to $1$. Our findings reveal that the standard occupancy model underestimates the zero-inflated probability (occupancy), and the bias increases as the community parameter $c$ increases. However, an interesting finding is that the zero-inflated N$-$mixture model can estimate the occupancy consistently as $c$ approaches $0$, but the bias can be positive or negative depending on the other parameters.

The paper is organized as follows: In Section 2, we develop the N$_c-$mixture model and present estimation methods. In Section 3, we consider the zero-inflated N$_c-$mixture model. Section 4 includes simulation studies to evaluate estimator performance. Two real data examples are presented in Section 5, and a discussion is in Section 6. Web Appendix A includes proofs for all propositions.

\section{N$_c-$mixture model}\label{section2}
\subsection{Notation and estimation}
Consider a multiple-visit sampling survey consisting of $n$ sites and $T$ visits. Let $N_{ij}$ be the number of individuals at site $i$ during the $j$-th visit. 
Following the motivation in the Introduction (see Figure \ref{FG1}),
we decompose $N_{ij}$ into two independent Poisson distributed random variables: $K_{i}$ and $M_{ij}$. The expected value of $K_{i}$ is $c\mu$ and the expected value of $M_{ij}$ is $(1-c)\mu$, where $0\le c \le 1$ and $\mu>0$ are constant parameters. Therefore, the number of individuals at each site, $N_{ij}$, follows an identical Poisson distribution with mean $\mu$, representing the abundance parameter over sites (per each visit). However, the numbers of individuals from multiple visits at a site, $N_{ij}, j=1,\cdots, T$, are not independent as they share a common variable $K_{i}$. The community parameter $c$ characterizes the degree of dependence between visits, as it also represents the correlation between each visit of $N_{ij}$. Note that $M_{ij}$ and $K_{i}$ are degenerate to 0 when $c=1$ and $c=0$, respectively.  
 
To model the data observation process, we assume that each individual is independently detectable with a probability of detection $r$. If the species is detected at site $i$ during visit $j$, let $Y_{ij}=1$, otherwise $Y_{ij}=0$. We denote the probability of detection at site $i$ during visit $j$ as $P(Y_{ij}=1)=p_{ij}$, then $p_{ij}=1-(1-r)^{N_{ij}}$ when $N_{ij}$ is known.  This forms an N$_c-$mixture model.

 It is clear that $Y_{ij}, j=1,\cdots, T$, are exchangeable variables, though not independent unless $c=0$. Let $Y_i = \sum_{j = 1}^T Y_{ij} $ be the frequency of occurrence at site $i$. In Web Appendix A, we show that the probability function of $Y_i$ under the N$_c-$mixture model is given by
 \begin{equation}\label{RNc-yi}
f(y_i;\bm\theta) = { T \choose y_i}  \exp (- d\mu r T  - c\mu)\sum_{k = 0}^{y_i} { y_i \choose k} (- 1)^k\exp \left\{c\mu {(1 - r)}^{T  - y_i + k} + d\mu r(y_i- k ) \right\},
\end{equation}
where  $\bm{\theta} = (\mu,r,c)$ is the vector of model parameters and $d=1-c$. 
Equation (\ref{RNc-yi}) is also referred to as the N$_c-$mixture model. 

When the community parameter $c=1$, the model in equation (\ref{RNc-yi}) simplifies to 
\begin{equation}\label{RN-yi}
f(y_i;\bm\theta_1) = { T \choose y_i}  \exp (-\mu)\sum_{k = 0}^{y_i} { y_i \choose k} (- 1)^k\exp \left\{\mu {(1 - r)}^{T  - y_i + k} \right\}, 
\end{equation}
where $ \bm{\theta}_1 =\bm{\theta}$ with $c=1$. 
This is the probability function of 
the N$-$mixture model \citep{royle2003}, but the explicit form (\ref{RN-yi}) is firstly given in \cite{haines2016a}. 

Similarly, when the community parameter $c=0$, the model (\ref{RNc-yi}) reduces to 
\begin{equation}\label{RNc-0}
f(y_i;\bm\theta_0) = { T \choose y_i} \left\{\exp (- \mu r)\right\}^{T  - y_i}\left\{1 -\exp(-\mu r)\right\}^{y_i},
\end{equation}
where $ \bm{\theta}_0 =\bm{\theta}$ with $c=0$.
The reduced model (\ref{RNc-0}) is a binomial distribution denoted as $Y_i \sim \text{Bino}(T, p)$, where $p=1-\exp(-\mu r)$. We also 
 note that the binary variables $Y_{ij}$ are independently and identically distributed with a mean of $E\{ 1-(1-r)^{N_{ij}}\}$ when $c=0$. A calculation using the Poisson moment generating function shows that $E\{ 1-(1-r)^{N_{ij}}\} =1- \exp(-\mu r) = p$, which leads to the same result as (\ref{RNc-0}). However, the model parameters $\mu$ and $r$ in (\ref{RNc-0}) cannot be separated, making the model unidentifiable, and only their product $\mu \times r$ is identifiable.

 When the community parameter $0<c<1$, the N$_c-$mixture model (\ref{RNc-yi}) is identifiable for $T\ge 3$. The log-likelihood function of the parameter vector $\bm{\theta}$ is given by $\ell(\bm{\theta}) = \sum\nolimits_{i = 1}^n \log \{f(y_i;\bm{\theta})\}$, and the maximum likelihood estimation is straightforward. For later use, the likelihood function can be represented in the framework of the multinomial model.
Let $m_j=\sum_{i=1}^n \rI(y_i=j)$ for $j=0,\cdots,T$, where $\rI(\cdot)$ is the indicator function. These statistics $m_j$ can be viewed as a result of the multinomial model with cell probabilities $ f( j;\bm\theta) $.
Therefore, we can write $\ell( \bm\theta  ) = \sum_{j = 0}^T  m_j\log \left\{ f( j;\bm\theta  ) \right\}$, and the score function for $\bm{\theta}$ is 
\begin{equation}\label{eqnscore}
S( \bm\theta  )  = \sum\limits_{j = 0}^T  \frac{\partial f(j;\bm\theta)}{\partial \bm\theta}\left\{ \frac{m_j - nf(j;\bm\theta )}{f(j;\bm\theta)} \right\}.
\end{equation}
\subsection{Occupancy rate}\label{section2.2}

\citet{royle2003} define the occupancy rate $\psi$ as a derived parameter under the N$-$mixture model. Specifically, $\psi = P(K_i>0)= 1-\exp(-\mu)$ in terms of our notations. Under the N$_c-$mixture framework, the occupancy rate per each visit can be defined as $P(N_{ij}>0)=1-\exp(-\mu)$. Alternatively, the rate can be defined as $1- P(N_{ij}=0,~\forall ~1\le j \le T)$, where if some $N_{ij}>0$, the site $i$ is considered occupied by the species. In this way, the occupancy rate is $1-\exp\left\{-\mu-(T-1)d\mu\right\}$, which depends on the number of visits $T$ and converges to $1$ when $T$ increases infinitely.
Both definitions, therefore, differ from the current concept of site occupancy. The problem is that the number of individuals at site $i$ may vary from visit to visit because $M_{ij},j=1,\ldots, T,$ are random in the N$_c-$mixture model. Therefore, rather than directly defining occupancy, we would like to include a zero-inflated parameter to determine site occupancy under the N$_c-$mixture model. This extension will be addressed in Section \ref{section3}.

\subsection{Behavior of the N$-$mixture model estimators} \label{section2.3}

In this subsection, we examine the behavior of N$-$mixture model estimators for $\mu$ and $r$ when the number of individuals at a site during multiple visits is not a fixed constant (i.e., $c<1$).

First, we examine the scenario where the community parameter $c$ is known and $T=2$ (double-visit). In this scenario, we use the notation $\tilde{\mu}_c$ and $\tilde{r}_c$ to represent the (restricted) maximum likelihood estimators (MLEs) of $\mu$ and $r$, respectively. 
 As shown in Web Lemma 1 of Web Appendix A, the MLEs have closed forms when $T=2$, which are given by $\tilde{\mu}_{c} = cz_1^2/(2z_1 + z_2)$ and $\tilde{r}_{c} = (2z_1 + z_2)/(cz_1)$, where $z_1 = \log \left\{ 2n/(2m_0+m_1 ) \right\}$ and $z_2 = \log (m_0/n)$. These closed forms allow us to easily determine the limits of the MLEs of the N$-$mixture model, $\tilde{\mu}_{1}$ and $\tilde{r}_{1}$, as given in the following proposition.

\begin{theorem}\label{Double-visit}
Under the N$_c-$mixture model with a double-visit survey, as the number of sites increases to infinity, the estimators $\tilde{\mu}_{1}$ and $\tilde{r}_{1}$ converge to $\mu/c$ and $cr$ respectively with probability one, for all $0<c\le 1$.
\end{theorem}

Proposition \ref{Double-visit} delivers insights into the behavior of N$-$mixture model estimators when the community parameter is less than $1$. The results show that the MLEs are consistent when $c=1$, which is a common property of the maximum likelihood approach.
 Additionally, the results indicate that the N$-$mixture model overestimates the abundance parameter $\mu$, and the bias increases as the community parameter $c$ decreases. In contrast,  the N$-$mixture model exhibits the opposite bias behavior for the detection probability $r$. Despite these biases, the estimator  $ \tilde{\mu}_{1} \times \tilde{r}_{1}$ can consistently estimate $\mu\times r$  under the framework of N$_c-$mixture model, which is noteworthy.

We initially believed that the results of Proposition \ref{Double-visit} would hold for surveys with more than two visits ($T>2$), but further investigation revealed that this is only partially true. The correct part is that there are moment estimators of the N$-$mixture model that exhibit similar behaviors to those described in Proposition \ref{Double-visit}.
To demonstrate this, we derived the moments of $Y_i$ under the N$-$mixture model and defined the resulting estimators as 
 $\tilde{\mu}_{1\rm{M}}$ and $\tilde{r}_{1\rm{M}}$. The following proposition summarizes the results of this analysis.

\begin{theorem}\label{CPMM}
The method of moment estimators of the N$-$mixture model, $\tilde{\mu}_{1\rm{M}}$ and $\tilde{r}_{1\rm{M}}$, are obtained by solving the equations
\begin{equation*}
    \label{moments}
\bar Y=Tp~~\mbox{and~} ~
\bar Y^2 =Tp +T(T-1)\{ 2p-1 +(1-p)^{2-r}\},
\end{equation*}
 where  $\bar Y$ is the sample average  and $\bar Y^2=\sum_{i=1}^n Y_i^2/n $.
 Furthermore, if the N$_c-$mixture model is true, as the number of sites increases to infinity, the estimators $\tilde{\mu}_{1\rm{M}}$ and $\tilde{r}_{1\rm{M}}$ converge to $\mu/c$ and $cr$ respectively with probability one for all $0<c\le 1$.
\end{theorem}

In the case of multiple-visit surveys, our simulation study (Section \ref{section4}) found that the limit of 
 the MLE of the N$-$mixture model, $\tilde{\mu}_{1}$, 
 exhibits a similar pattern to $\mu/c$ when $0<c <1$.  However, there is a discrepancy between them, and we can only determine the behavior  when the community parameter $c$ is close to $0$.

\begin{theorem}\label{Nmix-CP}
Under the N$_c-$mixture model with $c$ approaching zero, as the number of sites $n$ increases to infinity, the estimators $\tilde{\mu}_{1}$ and $\tilde{r}_{1}$ converge such that $\tilde{\mu}_{1}\rightarrow \infty$, $\tilde{r}_{1} \rightarrow 0$, and $\tilde{\mu}_{1} \tilde{r}_{1} \rightarrow \mu r$ with probability one.
\end{theorem}

Based on the results presented, we suggest that when $c<1$, the MLE of the N$-$mixture model tends to overestimate the parameter $\mu$ and underestimate the parameter $r$, and the bias increases as $c$ moves further away from $1$. However, the estimator $\tilde{\mu}_{1} \times \tilde{r}_{1}$ may still be able to consistently estimate $\mu\times r$ at certain ranges of $c$, such as when $c$ is close to 0, within the framework of the N$_c-$mixture model.

\section{Zero-inflated N$_c-$mixture model}\label{section3}

To account for the species occupancy, we extend the model (\ref{RNc-yi}) by incorporating a zero-inflation component.
Following \cite{mackenzie2002}, let $\psi$ be the site occupancy probability, then the probability of a zero count    ($Y_i=0$) is $(1-\psi)+ \psi f(0;\bm\theta)$. The likelihood function thus has an additional parameter $\psi$  so that we may write   
\begin{equation}\label{ZI-RNc}
L(\bm\theta ,\psi)  =\prod_{i=1}^n\left\{  (1-\psi)\rI(y_i=0) + \psi f(y_i;\bm\theta)\right \}.  
\end{equation}
We refer the model (\ref{ZI-RNc}) as a zero-inflated N$_c-$mixture (ZIN$_c$) model. 

When $c=1$, the model (\ref{ZI-RNc}) becomes a zero-inflated N$-$mixture (ZIN) model, as previously described in \cite{haines2016a}. In the case of $c=0$, it simplifies to a zero-inflated binomial (ZIB) model with the detection probability $p = 1 - \exp (- \mu r)$, which is also known as the first occupancy model \citep{mackenzie2002}.
Thus, the ZIN$_c$ model unifies both the ZIB and ZIN models into a single framework, with the special cases of $c=0$ and $c=1$ corresponding to ZIB and ZIN, respectively.

\subsection{Estimation}\label{Section3.1}
When the community parameter $0<c<1$,  the ZIN$_c$ model is identifiable for $T \geq 4$. 
By defining $p_0(\btheta, \psi)=(1-\psi)+ \psi f(0;\bm\theta)$ and $f(+;\bm\theta)={1- f(0;\bm\theta)}$, the likelihood function can be written as
$$
L(\bm\theta ,\psi)  = L_0(\bm\theta ,\psi) \times L_1(\bm\theta),
$$
where  $ L_0(\bm\theta ,\psi)=\{ p_0(\btheta, \psi)\}^{m_0}
\{ 1-p_0(\btheta, \psi)\}^{n-m_0}$ and  
$ 
L_1(\bm\theta)= \prod\limits_{j = 1}^T  
\left\{f( j;\bm\theta  ) /f( +;\bm\theta  ) \right\}^{m_j}.  $
Note that $L_0$ reflects the probability function of $\rI(Y_i>0)$ while $L_1$ is the conditional likelihood based on $Y_i>0$. As the conditional likelihood function is independent of the occupancy probability $\psi$, it allows us to estimate $\btheta$ without the confounding of $\psi$ \citep{KH2020}. Specifically, we can find the conditional MLE of $\btheta$, denoted as $\hat\btheta$, by solving the score function of $L_1(\btheta)$. Additionally, by taking the MLE of the profile likelihood $L_0(\hat\btheta ,\psi)$ we can find $\hat\psi=(n-m_0)/\{ nf(+;\hat\btheta)\} $.  As shown in Web Appendix A, Web Lemma~2 confirms that the estimators $\hat\btheta$ and $\hat\psi$ resulting from this method are also the usual MLEs based on (\ref{ZI-RNc}). The asymptotic variances of $\hat\btheta$ and $\hat\psi$ can be derived in the usual way.

In the zero-inflated type models, the main focus is on the estimation of the occupancy probability. Let $\tilde \psi_c$ be the MLE of $\psi$ for the ZIN$_c$ model,  given that the community parameter $c$ is known. 
We next examine the behavior of $\tilde \psi_0$ and $\tilde \psi_1$, which correspond to the occupancy estimators for the ZIB and ZIN models, respectively.

\subsection{Behaviors of $\tilde \psi_0$ and $\tilde \psi_1$}\label{section3.2}
Like the N$_c-$mixture model, the ZIN$_c$ model also has an identifiability issue for $\mu$ and $r$ when $c=0$. 
 As a result, under the ZIB model, only the parameter $p$ (i.e., $1-\exp(-\mu r)$) and the occupancy probability $\psi$ can be estimated. In practice, to fit ZIB models, it is common to set $r=1$ to estimate $\mu \times r$ with the resulting abundance estimator ($\tilde\mu_0$).

\begin{theorem}\label{ZIB-psi0}
Under the ZIN$_c$ model with $c>0$, the ZIB occupancy estimator $\tilde{\psi}_0$ shows an underestimation of $\psi$ with probability one as $n$ increases to infinity. 
A linear approximation of this underestimation can be represented as
$\tilde{\psi}_0\approx \frac{p}{p+\Delta} \psi$ where
\begin{equation}\label{rho}
\Delta  = \frac{\{1 - (1-p)^T\}\left\{ f(0;\bm\theta)-(1-p)^T  \right\}}{\left[ \frac{1}{p}\{1 - (1-p)^T\}-T (1-p)^{T -1} \right]\{ 1-f(0;\bm\theta)\}},  
\end{equation}
and $\Delta$ increases as $c$ increases, when $c=0$ then $\Delta=0$.
\end{theorem}
We notice that as either $T$ or $\mu \times r$ increases, the value of $\Delta$ decreases to zero. This result is expected, as when there are more visits or when the species abundance is high, the observed occupancy approaches $\psi$. The linear approximation bias provides a reasonable representation of the trend of $\tilde \psi_0$ in various aspects, depicting that as $c$ moves away from $0$ (or as the correlation between visits increases), the underestimation becomes more significant.

{\bf Remark~1.} As a direct consequence of Proposition \ref{ZIB-psi0}, we can also see that the corresponding abundance estimator $\tilde \mu_0$ increases as $c$ increases, and that $\tilde \mu_0$ at $c=0$ is a consistent estimator of the product of species abundance and detection probability ($\mu \times r$).

Estimators of the ZIN model tend to have bias when the community parameter $c$ is less than $1$. However, the behavior of these biases is complex. For example, the bias of the occupancy estimator $\tilde{\psi}_1$ does not vary monotonically with decreasing $c$. Despite this, when $c\approx 0$, the estimators of the ZIN model (except $\tilde{\psi}_1$) behave similarly to those of the N$-$mixture model, as shown in Proposition \ref{Nmix-CP}. Interestingly, $\tilde{\psi}_1$ can consistently estimate $\psi$ at $c=0$, as shown in the next proposition. For clarity,  if there is no confusion, $\tilde \mu_1$ and $\tilde r_1$ in this proposition also refer to the MLE of the ZIN model (or the restricted MLE of the ZIN$_c$ model).

\begin{theorem}\label{ZINc-CP}
Under the N$_c-$mixture model with $c$ approaching zero, as the number of sites $n$ increases, the estimators $\tilde{\mu}_{1}$ and $\tilde{r}_{1}$ converge such that $\tilde{\mu}_{1}\rightarrow \infty$, $\tilde{r}_{1} \rightarrow 0$, and $\tilde{\mu}_{1} \tilde{r}_{1} \rightarrow \mu r$ with probability one. Additionally, $\tilde{\psi}_1$ is a consistent estimator for $\psi$ when $c=0$.
\end{theorem}
The estimator $\tilde{\psi}_1$ is also a consistent estimator under the ZIN (or ZIN$_c$ with $c=1$) model. Therefore, Proposition \ref{ZINc-CP} shows that $\tilde{\psi}_1$ behaves like a bridge with both ends (at $c=0$ and 1) at the same level; however, it is not clear whether the bridge deck is always above or below this level, or if it varies in different sections. We will further investigate this behavior through a simulation study.

\subsection{Tests for ZIN and ZIB}\label{section3.3}
The null hypothesis of the ZIN or ZIB model can be tested within the framework of the ZIN$_c$ model, as it is equivalent to testing the values $\{c=1\}$ or $\{c=0 \}$. A simple method to justify the hypothesis is to use a Wald-type confidence interval based on the estimate of $c$. A more formal approach is to use the likelihood ratio test, as the null hypothesis is a submodel of the full ZIN$_c$ model. However, it is important to note that the asymptotic distribution of the likelihood ratio test under the null hypothesis is a mixture of $0$ and chi-square distributions, rather than the usual chi-square distribution \citep{selfliang1987}. In practice, we also suggest generating bootstrap samples under the null hypothesis to find the $p-$value. This is similar to the conclusion of \cite{dail}.

\section{Simulation study}\label{section4}

We conducted simulations to evaluate the performance of proposed models and estimators. We considered  two scenarios:
the N$_c-$mixture model and the ZIN$_c$ model.  In the first scenario, we computed the maximum likelihood estimators for both the N$-$mixture model and N$_c-$mixture model. In the second scenario, we calculated the maximum likelihood estimators for the ZIB, ZIN, and ZIN$_c$ models. All estimates were calculated by using the \texttt{optim} function in the  \texttt{R} software \citep{rct2022}.

 We specified the true parameter values as $\mu =1, 2$, $r=0.25,0.5$, and $c=0,0.05,\ldots,1$ for the simulations. For both scenarios, the number of sites was set at $n=200,500,1000$, and the number of visits was set at $T=5,7,10$. In the second scenario, we also set the occupancy probability to $\psi=0.7$. We generated 1,000 data sets for each parameter setting and calculated the estimates and associated standard error estimates for each data set.

 To account for outliers in some of the simulated data sets, we present the median of the parameter estimates (Med), the median of the estimated standard error (Med.se), and the median absolute deviation (MAD) scaled to align with the normal distribution. Additionally, we report the coverage percentage (CP) of the nominal $95\%$ Wald-type confidence intervals. 
We note that in some instances, the numerical methods utilized for estimating the parameters for each model did not converge, with a higher frequency of non-convergence observed for the ZIN model and a lower frequency for the ZIN$_c$ model. However, in most cases, the percentage of failures was minor and not reported.

\subsection{Simulation study A: N$_c-$mixture model}\label{section4.1}
\begin{figure}[h!]
\begin{center}
\includegraphics[width=5.8in]{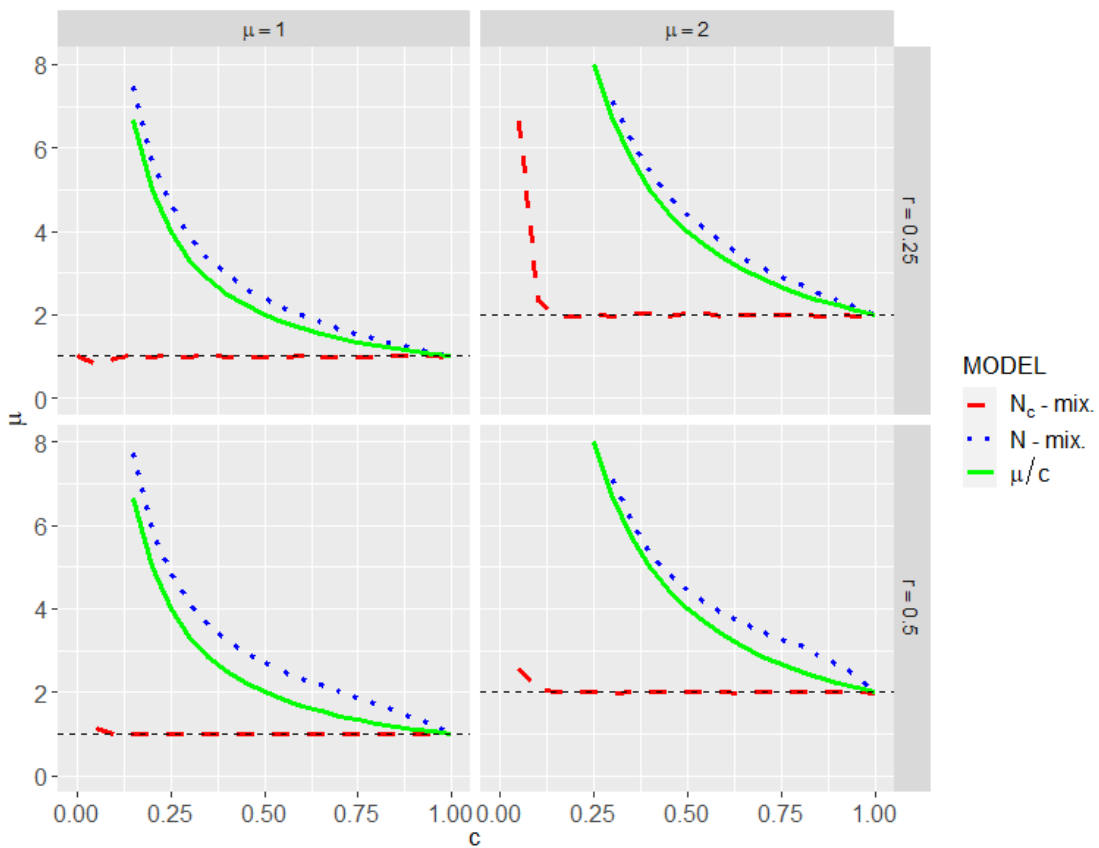}
\end{center}
\caption{
Median estimates of abundance $\mu$ for N$-$mixture and N$_c-$mixture models (Simulation study~A) as a function of the community parameter $c$, with the number of visits $T=7$ and sites $n=500$. 
The sub-graphs correspond to four combinations of $\mu=1,2$ and individual detection probabilities $r=0.25,0.5$. Data points with high values were removed for clarity.
}\label{SSA-Fig}
\end{figure}

The median estimates of $\mu$ for $T=7$ and $n=500$ are displayed in Figure~\ref{SSA-Fig}. A comprehensive examination of the simulation results for $c=(0.25,0.5,0.75)$ can be found in Web Tables 1-6 in Web Appendix B.

Figure \ref{SSA-Fig} illustrates the behavior of the N$-$mixture abundance estimator $\tilde \mu_1$ as a function of the community parameter $c$. The figure shows that the estimator exhibits a positive bias for all values of $c$ less than $1$. This bias follows a monotonically decreasing pattern and only approaches the true value when $c$ equals $1$. When $c$ is close to $0$, the bias can be substantial but has been removed from the figure for ease of visualization. Additionally, the figure shows that $\tilde \mu_1$ is greater than $\mu/c$  for all values of $c$ less than 1, although the difference has not been explicitly explored. In a separate simulation study (not reported), we calculated the moment estimator for the N$-$mixture  model and found that its pattern closely resembled the reference curve $\mu/c$. These simulation results agree  with the findings outlined in Propositions 1-3. Corresponding  conclusions can also be drawn from the results of the estimator $\tilde r_1$ shown in Web Figure 1 of Web Appendix B.
The MLE of the N$_c-$mixture model can consistently estimate $\mu$, but it also frequently  exhibits bias around $c=0$.
This is mainly because the parameters of $\mu$ and $r$ are almost unidentifiable when $c\approx 0$. The bias is more pronounced when $\mu$ is large and $r$ is small, but it becomes smaller when $n$ or $T$ increases; see Web Table~6 (for $n=1000$ and $T=10$ cases).

In Web Tables 1-6, it can be seen that the N$-$mixture abundance estimator has a relative bias ranging from $40\%$ to $400\%$ when $c$ varies from $0.75$ to $0.25$. The bias is more pronounced when $T$ or $r$ increases but less severe when $\mu$ increases. On the other hand, the N$_c-$mixture model shows nearly unbiased estimates in all cases, except when $n = 200,~ T = 5$, and $r = 0.25$, where the relative bias can reach up to $7.5\%$-$16\%$ when $\mu$ varies from $2$ to $1$; see Web Tables 1 and 4.

In terms of mean absolute deviation (MAD), both models show a decrease in variation as $c$ increases, with the N$-$mixture model showing a much steeper decrease compared to the N$_c-$mixture model. Specifically, when $c=0.25$, the MAD of $\tilde\mu_1$ can be $2$ to $4$ times that of $\hat\mu$, but the former is usually smaller than the latter when $c=0.75$. The asymptotic standard error estimates, as measured by Med.se, generally match the results of the corresponding MAD, making them reasonably reliable for the scenarios considered. The Wald-type confidence interval estimator of the N$_c-$mixture model performs well when data information is sufficient, with a close match to the nominal $95\%$ confidence level at $n\ge 500$ and $T\ge 7$. However, in some cases with small $r$ and $c$, the coverage probability (CP) can be lower than $80\%$ (as seen in Web Tables 1, 4, and 5), indicating an unsatisfactory performance. The N$-$mixture model estimator often reaches $0$ for the coverage probability due to the severe bias problem of $\tilde \mu_1$ when $c \leq 0.75$. 

The results from Web Tables 1-6 on the detection probability parameter $r$ are similar to $\mu$, with the only difference being that the N$-$mixture model estimator's bias is in the opposite direction. Finally, it is found that the median of the product $\tilde \mu_1 \times \tilde r_1$ presents nearly unbiased results for estimating $\mu\times r$ in all cases. Note that this property is only proven for the range $c\approx 0$ (Proposition 5), but the simulation results suggest that the range of $c$ can be extended somewhere.

\subsection{Simulation study B:  ZIN$_c$ model}\label{section4.2}
In Figures 3 and 4, we present the median estimate results for the parameters $\mu$ and $\psi$ when $T=7$ and $n=500$. The detailed simulation results for $c=(0.25,0.5,0.75)$  can be found in Web Tables 7-12 of the Web Appendix B. In the ZIB model, we have fixed the value of $r = 1$. This is because the parameter $\mu \times r$ is non-separable in this case. 
 
\begin{figure}[ht!]
\begin{center}
\includegraphics[width=5.8in]{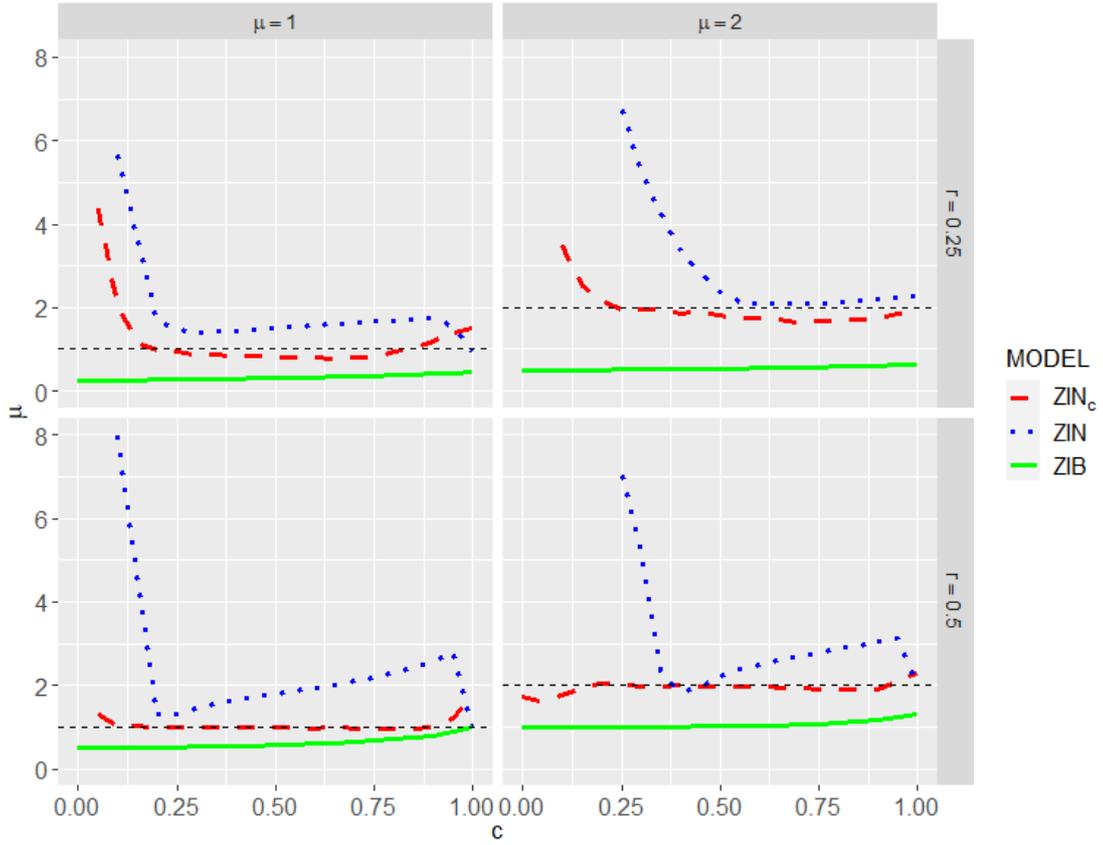}
\end{center}
\caption{
Median estimates of abundance $\mu$ for ZIB, ZIN$_c$, and ZIN models (Simulation study~B) as a function of the community parameter $c$, with the number of visits $T=7$ and sites $n=500$. 
The sub-graphs correspond to four combinations of $\mu=1,2$ and individual detection probabilities $r=0.25,0.5$. 
Data points with high values were removed for clarity.
}\label{SSB-mu}
\end{figure}

In Figure \ref{SSB-mu}, we can see that the ZIB estimator $\tilde \mu_0$ consistently underestimates $\mu$. In fact, $\tilde \mu_0$ is approximately equal to $\mu\times r$ when $c$ is close to 0 and increases as $c$ increases (as stated in Remark 1). The ZIN estimator $\tilde \mu_1$ has a similar trend as the N$-$mixture model when $c$ is close to $0$, but it falls off quickly, rises slowly, and becomes consistent with $\mu$ at $c=1$.  In general, $\tilde \mu_1$ mainly exhibits positive bias but occasionally also shows negative bias at some $0<c<1$. The ZIN$_c$ estimator $\hat \mu$ typically exhibits some bias at both ends of the range of $c$, which can be very large at $c\approx 0$ when $r$ is small.  As expected, increasing $n$ and $T$ can mitigate bias issues of the maximum likelihood estimator.

\begin{figure}[ht!]
\begin{center}
\includegraphics[width=5.8in]{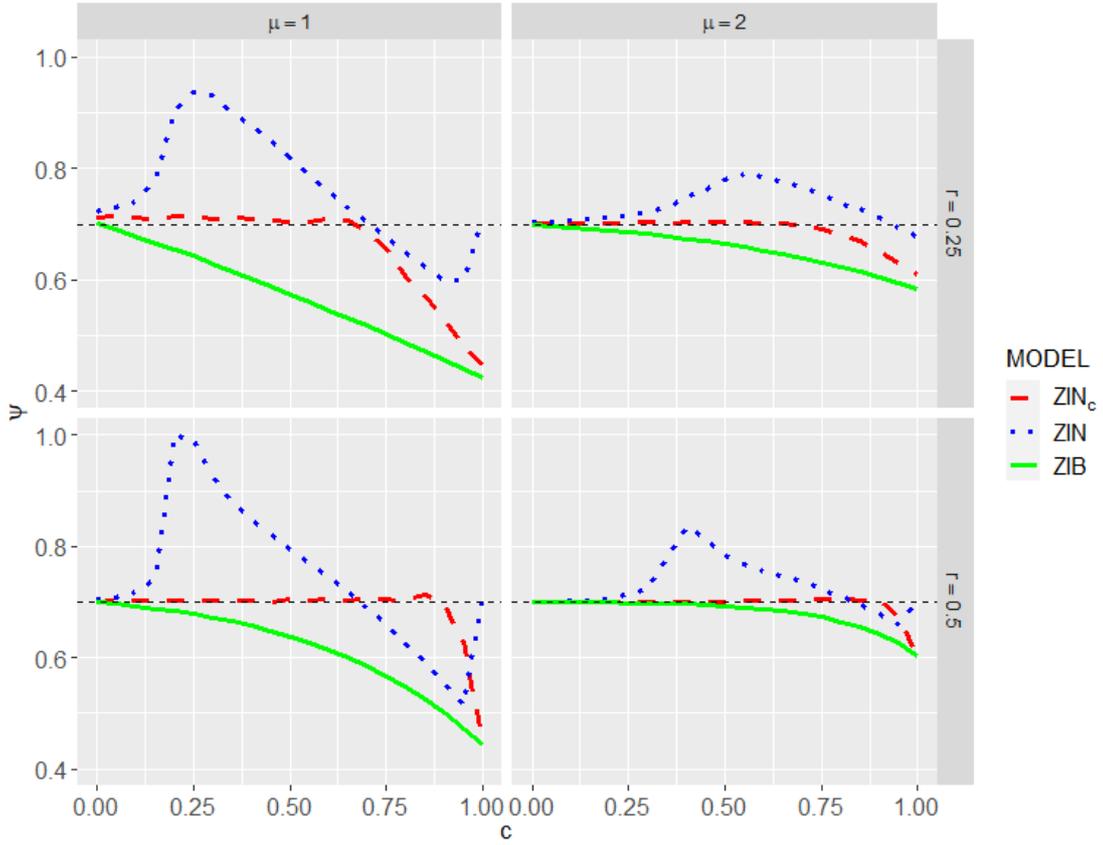}
\end{center}
\caption{Median estimates of occupancy rate $\psi$ for ZIB, ZIN$_c$ and ZIN models  (Simulation study~B) plotted against the community parameter $c$, with  the number of visits $T=7$ and sites $n=500$. 
The sub-graphs correspond to four combinations of $\mu=1,2$ and individual detection probabilities $r=0.25,0.5$.
}\label{SSB-psi}
\end{figure}

In Figure \ref{SSB-psi}, the ZIB estimator $\tilde \psi_0$ was found to underestimate $\psi$ when $c>0$, with a greater bias observed when $\mu=1$ compared to $\mu=2$. The relative bias reached a maximum of $-35\%$ when $c=1$ and $\mu=1$. The approximate formula of Proposition~4 was found to be consistent with the behavior trend of $\tilde \psi_0$, although it is not shown in the figure.
The ZIN occupancy estimator $\tilde \psi_1$ was found to fit well with $\psi$ at both ends of the $c$ range, as stated in Proposition 5. A positive bias was observed around $c\approx 0$ and a negative bias around $c \approx 1$. The partial derivative of $\frac{\partial}{\partial c}\tilde \psi_1$ was proven to be positive when $c\approx 0$, indicating overestimation of $\psi$ in that region. However, at $c\approx 1$, the partial derivative was found to be positive in most situations, with some negative signs observed in some instances, such as when $\mu=2$ and $r=0.15$. In this case, the plot of $\tilde \psi_1$ against $c$ exhibited behavior similar to an arch bridge, with only positive biases observed for all $0<c<1$.
The ZIN$_c$ estimator $\hat \psi$ was found to be unbiased in general, except when $c\approx 1$. The bias was found to be smaller when $\mu$ or $r$ is increased.

Next, we present a more detailed summary of the performance of the three occupancy estimators based on the results in Web Tables 7-12. 
The upper half of Web Table~7 ($\mu \times r=0.25$ and $T=5$)  shows that the ZIB occupancy estimator $\tilde{\psi}_0$ has a relative bias of $-11\%$ at $c=0.25$, which becomes more pronounced as $c$ increases, reaching a maximum of $-31\%$ at $c=0.75$. However, increasing $T$ or $\mu \times r$ can reduce the bias of $\tilde{\psi}_0$ as seen in Web Tables 7-12, which is consistent with Proposition \ref{ZIB-psi0}.
In Web Tables 7-9 ($\mu=1$), the ZIN occupancy estimator $\tilde{\psi}_1$ has a relative bias of around $28\%$-$40\%$ at $c=0.25$. Specifically, the ZIN model often estimated $\psi$ as one at $c=0.25$ when $T\ge 7$ and $r=0.5$ (Web Tables 8-9). The bias of $\tilde{\psi}_1$ decreases as $c$ increases, showing a small negative bias at $c=0.75$. When $\mu$ is increased to $\mu=2$, Web Tables~10-12 show that $\tilde{\psi}_1$ overestimates $\psi$ in all cases, with the most significant bias around $10\%$-$13\%$ at $c=0.5$, and small or even negligible bias at $c=0.25, 0.75$.
 Web Tables 7-12 also reveal that the bias of the ZIN$_c$ occupancy estimator $\hat{\psi}$ is generally close to $0$, except for a few cases of $r=0.25$, $T=5$, and $c=0.75$ (Web Table 7).

The ZIB occupancy estimator has the lowest MAD, indicating that it is the most stable of the three methods.  When $\mu=1$, the MAD of the ZIN estimator decreases with increasing $c$; however, when $\mu=2$, the MAD is more prominent at $c = 0.5$ due to the significant bias of the ZIN estimator at this point. In contrast,  the MAD of the ZIN$_c$ estimator increases with increasing $c$ values; when $\mu=1$,  the MAD at $c = 0.75$ can be $1.5$ to $3$ times the MAD at $c = 0.25$, but the increase is much smaller when $\mu=2$.  Generally, the stabilities of all three estimators improve with increasing values of $n,~ T$, $r$, and $\mu$.

The Med.se of the ZIB occupancy estimator fits the corresponding MAD reasonably well, except for some negative biases observed at $c=0.75$. 
However, the resulting interval estimator is only reliable when $\mu=2$ and $c=0.25$; in other cases, the corresponding CP may even reach zero showing  very poor performance.   
For the ZIN occupancy estimator, Med.se often overestimates MAD, particularly with increasing $c$, but this is less of an issue when $r$ or $\mu$ are increased. The resulting CP also shows many abnormal values, even when it appears to be close to the nominal level in some cases. For example, Web Tables 8-9 show that the CP can reach about $99\%$ in the upper panels but may drop to zero in the bottom panels. As a result, the interval estimates for the ZIN model are not considered reliable.
The ZIN$_c$ estimator, being a maximum likelihood estimate, can perform well on Med.se and CP measures when the data information is sufficient, but its performance is not guaranteed otherwise.

Lastly, we note  that abundance estimators can be similarly affected by issues such as violation of model assumptions or a lack of sufficient data, which can result in substantial bias in some instances, such as when $c=0.25$ and $r=0.25$. 
In general, compared to abundance estimators, the corresponding occupancy estimators are relatively less affected by these issues and their performance is relatively robust.

\section{Examples}\label{section5}
\subsection{Example~1. Fisher data}\label{SectionFishers}
The fisher $(Martes~ pennanti)$ 
is a carnivorous mammal native to the boreal forests of North America. In 2000, a survey program using noninvasive methods was conducted to collect data on fisher species distribution in northern and central California \citep{zielinski2005,royledorazio2008}. 
The data consists of multiple-visit occurrences at 
$n = 464$ sites, each visited $T = 8$ times. See Web Table~13 for the data. Out of the eight visits, $400$ sites had zero counts, resulting in a sample occupancy rate of $64/464 = 13.8\%$. We analyzed the fisher data using the models presented in Sections \ref{section2} and \ref{section3}. The results are summarized in the top panel of Table 1, where we also calculated the Akaike information criterion (AIC) value for each model to evaluate model fit.

The AIC values for the ZIN and ZIN$_c$ models were significantly lower than those of the N$-$mixture and N$_c-$mixture models, respectively, indicating that including zero-inflated probabilities improves the fit of the model to the fisher data. Among the three zero-inflated models, the ZIN$_c$ had the smallest AIC value. Furthermore, the ZIN$_c$ model estimated the community parameter $c$ to be in the middle of $(0, 1)$ and the associated Wald-type confidence interval (Web Table~14) provided strong evidence against the null hypotheses $c = 0$ and $c = 1$. The likelihood ratio statistics for the tests $c = 0$ and $c = 1$ were $40.21$ and $8.23$ respectively, and the $p-$values based on 10,000 bootstrap samples were $0$ and $0.006$, respectively, leading to the same conclusion for rejecting ZIB and ZIN models. To sum up, the ZIN$_c$ model fits the fisher data better than the other models, however, the occupancy estimates for the three zero-inflated models are similar.

It is worth noting that the ZIN model estimated $\psi$ to be $0.175$, which appears larger than the other models, but the ZIN model has an unoccupied probability of $(1-\psi)+ \psi\exp(-\mu) $. Therefore, to compare occupancy rates with other models, we should use $\psi \{1-\exp(-\mu)\}$ with an estimate of $0.175 \{1-\exp(-1.79)\}=0.146$ (called the occupied probability of the ZIN model), which is similar to other estimates of $\psi$. In contrast, the resulting abundance estimates are quite different, with ZIN reporting a higher value and ZIB having the lowest value. The standard error estimates for the three models also follow the same order, with the parameter estimates for ZIN showing the greatest estimated variation. This phenomenon is similar to what was observed in Web Table 8 of Simulation Study B.

\newpage
\begin{table}[ht]\footnotesize
\centering
\begin{tabular}{llllll}
   \hline
Model & $\mu$ &  $r$ & $c$ &  $\psi$  & AIC \\ 
\hline
  &\multicolumn{5}{c}{\textbf{Fisher}}\\
 \cline{3-5}
ZIN$_c$& 0.93 (0.21)& 0.49 (0.10)& 0.51 (0.13)& 0.154   (0.020)& 626.9 \\
  ZIN & 1.79  (0.68)& 0.22 (0.05)   & $1^{*}$ & 0.175   (0.030)& 633.2\\
  ZIB &  0.51  (0.04) & $1^{*}$ & $0^{*}$ & 0.140 (0.016)& 663.1 \\ 
 N$_c-$mix. & 0.13 (0.02)& 0.46 (0.03)& 0.93 (0.02) &  & 636.8 \\ 
  N$-$mix. & 0.16 (0.02)& 0.37 (0.02)& $1^{*}$ & 0.150 (0.017)& 650.6 \\ 
   \hline

&\multicolumn{5}{c}{\textbf{Blue Jay}}\\  
  \cline{3-5}
  
  ZIN & 48.96 (427.97) & 0 (0.04) & $1^{*}$ & 0.729 (0.091) & 170.1 \\ 
  ZIB & 0.22 (0.03) & $1^{*}$ & $0^{*}$ & 0.723 (0.077) & 168.1 \\ 
  N$-$mix. & 1.64 (0.56) & 0.09 (0.03) & $1^{*}$ & 0.806 (0.109) & 169.5 \\
   \hline
   
	&\multicolumn{5}{c}{\textbf{Catbird}}\\
  \cline{3-5}
ZIN$_c$ & 0.60 (0.32) & 0.42 (0.21) & 0.19 (0.14) & 0.421 (0.079) & 139.1 \\ 
  ZIN & 0.85 (1.31) & 0.16 (0.09) & $1^{*}$ & 0.740 (0.691) & 137.8 \\ 
  ZIB & 0.26 (0.04) & $1^{*}$ & $0^{*}$ & 0.403 (0.074) & 138.5 \\ 
  N$_c-$mix. & 0.52 (0.17) & 0.19 (0.06) & 0.97 (0.12) &  & 137.9 \\ 
  N$-$mix. & 0.55 (0.14) & 0.18 (0.04) & $1^{*}$ & 0.421 (0.079) & 135.9 \\ 
   \hline

 &\multicolumn{5}{c}{\textbf{Common Yellow-Throat}}\\
     \cline{3-5}
ZIN$_c$& 0.93 (0.23)& 0.47 (0.09) & 0.56  (0.10)& 0.775   (0.075)& 222.9 \\
  ZIN & 2.06 (0.81) &0.20 (0.05)  & $1^{*}$  & 0.851 (0.123) &227.8 \\
  ZIB & 0.49 (0.04) & $1^{*}$ &    $0^{*}$   & 0.723 (0.064)& 254.5 \\ 
  N$_c-$mix. & 1.09 (0.38)& 0.31 (0.09)& 0.88 (0.15)& & 226.7 \\
  N$-$mix.&1.45 (0.25)& 0.23 (0.03)  &$1^{*}$   &0.765 (0.059)& 226.8\\
  \hline
   
 &\multicolumn{5}{c}{\textbf{Tree Swallow}}\\ 
  \cline{3-5}
ZIN$_c$ & 0.75 (0.22) & 0.45 (0.08) & 0.61 (0.12) & 0.682 (0.106) & 196.4 \\ 
  ZIN & 1.81 (0.75) & 0.18 (0.05) & $1^{*}$ & 0.726  (0.135) & 204.0 \\ 
  ZIB & 0.40 (0.04) & $1^{*}$ & $0^{*}$ & 0.587 (0.071) & 228.0 \\ 
  N$_c-$mix. & 0.48 (0.11) & 0.46 (0.07) & 0.72 (0.08) &  & 198.4 \\ 
  N$-$mix. & 1.02 (0.19) & 0.22 (0.03) & $1^{*}$ & 0.641 (0.069) & 203.9 \\
   \hline 

 &\multicolumn{5}{c}{\textbf{Song Sparrow}}\\
 \cline{3-5}
 ZIN$_c$ & 1.01 (0.59) & 0.37 (0.08) & 0.90 (0.09) & 0.678 (0.240) & 184.7 \\ 
  ZIN & 2.23 (0.98) & 0.21 (0.06) & $1^{*}$ & 0.573 (0.105) & 187.9 \\ 
  ZIB & 0.54 (0.05) & $1^{*}$ & $0^{*}$ & 0.501 (0.072) & 210.5 \\ 
  N$_c-$mix. & 0.58 (0.13) & 0.42 (0.05) & 0.93 (0.04) &  & 183.4 \\ 
  N$-$mix. & 0.81 (0.16) & 0.31 (0.04) & $1^{*}$ & 0.555 (0.071) & 190.5 \\ 
   \hline 
   
\end{tabular}
\caption{
Parameter estimates (standard errors in
parenthesis) and AIC values for each model fitted to the fisher data and the five bird species data of the BBS survey. Note that $^{*}$ indicates that the parameter value is a fixed constant without estimation. The occupancy rate $\psi$ of  the N$-$mixture model is a deriving parameter defined as $\psi=1-\exp(-\mu)$; there is  no occupancy rate for the  N$_c-$mixture model (Section 2.2). For the blue jay data, the results for the ZIN$_c$ and N$_c-$mixture  models are omitted because they are reduced to ZIB and N$-$mixture models, respectively. 
The five bird species are sorted  according to the estimate of $c$ from the ZIN$_c$ model.
 }
\end{table}

\subsection{Example~2: Breeding Bird Survey (BBS) data}
This example aims to understand how the community parameter $c$ may vary among different species in data from the same survey. We consider the data used in \citet{royle2006}, originally from the North American Bird Breeding Survey (BBS) program, which includes records of occurrence data for five species of birds -- blue jay ($Cyanocitta~cristata$), catbird ($Dumetella~carolinensis$), common yellow-throat  ($Geothlypis~trichas$), tree swallow ($Tachycineta~bicolor$), and song sparrow ($Melospiza~melodia$)-- from $50$ locations with $11$ repeated visits. The sample occupancy rates for the five species (in the above order) are $66\%$, $38\%$, $72\%$, $58\%$, and $52\%$; see Web Table~13 for the data.

Results are reported in Table 1, where the order of the species was actually sorted according to the estimated value of $c$ in the ZIN$_c$  model. The first two species show small estimates of $c$, with the ZIN$_c$ model even degenerating to the ZIB model for the blue jay data due to an estimate of $c = 0$. All the considered models showed comparable AIC values for both species, indicating a similar level of model fit. However, the parameter estimates of $\mu,~r$, and $c$  for each model produced conflicting results. We believe this is a problem of model identifiability, similar to the one described \citet{royle2006} and \citet{link2003}. \citet{royle2006}  concluded that this problem is more pronounced in occupancy models when the probability of detection is low, which is the case for these two species.

The next two species, common yellow-throat and tree sparrow, show middle estimates of $c$, where both $95\%$ Wald type confidence interval does not include $0$ or $1$ (Web Table 14 ). The results for model comparisons here can be summarized as similar to Example~1, particularly for zero-inflated models. In terms of AIC, the ZIN$_c$ model performs better than other models. 

The last species, song sparrow, has a high estimate of $c$ of $0.90$, which is close to $1$. In fact, the upper limit of its $95\%$ confidence interval exceeds the upper bound of $1$. 
In this case, the  standard errors of $\hat\mu$ and $\hat\psi$ for the ZIN$_c$ model are relatively high, indicating that estimation uncertainty increases as $c$ approaches $1$. This phenomenon was also observed in Simulation Study B. 

In Web Table 14, we also report bootstrap $p-$values for the likelihood ratio tests to the hypotheses of ZIB and ZIN models, which suggest significant evidence against the null hypotheses for the last three species of BBS data.

\section{Discussion}\label{section6}
The N$-$mixture model has been extended to allow for the number of individuals at a sample site to vary with each visit. In this extension, the number of individuals at a site per visit is treated as a latent variable that is decomposed into two components, with only one of these components being considered in the original N$-$mixture model.
It should be noted that the extended N$_c-$mixture model is only applicable to closed populations (single season) due to the assumption of constant $\mu$. While it is possible to relax this assumption using regression models with the site and/or visit covariates, it is unclear whether this approach is effective for some multiseasonal data. 

Unlike the N$-$mixture model, the occupancy of the N$_c-$mixture model may not be well defined by the abundance parameter. 
To address this, we propose a zero-inflated N$_c-$mixture model that uses a zero-inflated probability to define occupancy explicitly. 
This new model unifies the commonly used standard occupancy and zero-inflated N$-$mixture models into one framework. As a result, we discover interesting properties of both models, such as the use of standard occupancy estimators under the zero-inflated N$-$mixture model or vice versa.

Our extension models offer greater flexibility in fitting data, but they also increase the complexity of the model. These models may also experience issues such as numerical instability, parameter identifiability, and model identification problems, 
particularly when data is sparse, such as in cases of low detection probability, low abundance, few replicates of visits, or a small sample of sites. These issues have been observed in both simulation studies and data analysis. Indeed, these problems were initially raised in N$-$mixture models \citep{dennis2015,barker2018,link2018} as a caution against their use. Our extension models require an additional community parameter, which may exacerbate these problems. However, these issues may be mitigated by  incorporating additional capture-recapture data \citep{barker2018}  or including covariates in the models \citep{link2018}. Further research is necessary to improve the estimation in these contexts.
  
There are several potential avenues for extending this work, each of which would require further implementation and investigation. 
\begin{itemize}
\item Using the negative binomial distribution to model abundance instead of the Poisson distribution. This distribution is often more appropriate for biological and ecological data analysis, but it also includes an additional aggregation parameter, increasing model complexity and estimation difficulty.
\item Examining count data instead of occurrence data. This would require a likelihood function without an explicit form, making estimation computationally intensive, particularly for large datasets.
\item Incorporating covariates into the models. Including covariates is important for real-world applications. This can be done by using a log link for abundance and a logit link for detection and occupancy probability. Incorporating species covariates for the community parameter in a joint species model may also be meaningful.
\item Extending to time-to-detection data. Time-to-detection occupancy models have been developed recently, which also show some advantages over occurrence data models \citep{priyadarshani2022}. \citet{strebel2021} propose an N$-$mixture time-to-detection occupancy model that enables estimating the abundance without marking individuals. An analogous N$_c-$mixture occupancy model for time-to-detection data is currently under development.

\end{itemize}



\section*{Acknowledgements}
This work was supported by the National Science and Technology Council  of Taiwan.


\end{document}